\documentclass[letter]{aa}  
\usepackage[flushleft]{threeparttable}
\usepackage{float}
\usepackage{graphics}
\usepackage{graphicx}
\usepackage{amssymb, amsmath}
\usepackage{tikz}
\usetikzlibrary{graphs, positioning, quotes, shapes.geometric}
\usepackage{amstext}
\usepackage[T1]{fontenc}
\usepackage{mathrsfs}
\usepackage[normalem]{ulem}
\usepackage{natbib,twoopt}
\usepackage{hyperref} 
\usepackage{soul,xcolor}
\usepackage{booktabs}
\usepackage{rotating}
\usepackage{txfonts}
\makeatletter
\def\@to{to}
\def\as     {\ifmmode {\rlap.}$\,$''$\,$\! \else ${\rlap.}$\,$''$\,$\!$\fi}
     \def\decsec  {\ifmmode {\rlap.}$\,$^{\rm s}$\,$\! \else ${\rlap.}$\,$^{\rm s}$\,$\!$\fi}\def\decss  {\ifmmode {\rlap.}$\,$^{\rm s}$\,$\! \else ${\rlap.}$\,$^{\rm s}$\,$\!$\fi}
\makeatother

\usepackage{amsmath}          
\usepackage{dcolumn}          
\usepackage{natbib}           
\usepackage{graphicx}         
\usepackage{txfonts}          
\usepackage{mathrsfs}         
\usepackage{rotating}         
\usepackage{subfigure}        
\usepackage{lscape}           
\usepackage{afterpage}        
\usepackage{xspace}           

\bibpunct{(}{)}{;}{a}{}{,}    
\newcolumntype{.}{D{.}{.}{-1}}
\newcolumntype{d}[1]{D{.}{.}{#1}}

\newcommand{\mrm}[1]{\ensuremath{\mathrm{#1}}}

\providecommand*{\diff}{\ensuremath{\mathrm{d}}}
\providecommand*{\ee}{\ensuremath{\mathrm{e}}}

\begin{document}

   \title{First detection of CHD$_{2}$OH towards pre-stellar cores}


   \author{Y. Lin         
          \inst{1}
          \and S. Spezzano
          \inst{1}
          \and P. Caselli\inst{1}}

   \institute{Max-Planck-Institut f{\"u}r Extraterrestrische Physik, Giessenbachstr. 1, D-85748 Garching bei M{\"u}nchen\\
              \email{ylin@mpe.mpg.de}
             }

   \date{Received ; accepted }

 
  \abstract
  {The inheritance of material across the star and planet formation process is traced by deuterium fractionation. We report here the first detection of doubly deuterated methanol towards pre-stellar cores. We study the deuterium fractionation of methanol, CH$_{3}$OH, towards two starless and two pre-stellar cores. We derive a D/H ratio of 0.8-1.9$\%$ with CH$_{2}$DOH in pre-stellar cores H-MM1 and L694-2, consistent with measurements in more evolved Class 0/I objects and comet 67P/Churyumov-Gerasimenko, suggesting a direct chemical link arising in the pre-stellar stage. Furthermore, the column density ratios of CHD$_{2}$OH/CH$_{2}$DOH are $\sim$50-80$\%$, consistently high as that towards Class 0/I objects, indicating an efficient formation mechanism of CHD$_{2}$OH, possibly through H atom additions to D$_{2}$CO. The CH$_{2}$DOH/CH$_{3}$OH and CHD$_{2}$OH/CH$_{3}$OH column density ratios in the two pre-stellar cores are larger than that in the two starless cores B68 and L1521E, representing an evolutionary trend of methanol deuteration in early-stage cores.}

   \maketitle
%

\section{Introduction}


High levels of deuterium fraction is an extraordinary chemical feature of dense cores, the precursors of low-mass stars and planetary systems. While the cosmic atomic D/H elemental ratio is 1.6$\times$10$^{-5}$ (\citealt{Linsky03}), the deuterium fraction (molecular D/H ratio) in dense, cold cores can be orders of magnitude larger (\citealt{Ceccarelli07}). The enrichment of molecules in deuterium is favored by low temperatures and high CO-depletion environment (e.g., \citealt{DL84}, \citealt{Walmsley04}, \citealt{Roueff05}). Molecules detected in dense cores have a wide range of D/H, which is reflective of the different formation times and mechanisms of different molecular species (\citealt{Taquet14}, \citealt{CC14}). Amongst all molecules, methanol (CH$_{3}$OH) shows a high D/H ratio of up to tens of percent  (\citealt{Parise06}). As CH$_{3}$OH is a critical molecule in advancing the chemical complexity of star-forming regions (\citealt{GH06}, \citealt{CC12}), measurements of the deuterium fraction provides important clues on its formation pathways and chemical evolution over the course of star formation.   

The formation of CH$_{3}$OH takes place mainly on the surface of icy grains mantles, through successive addition of H atoms to CO and H$_{2}$CO, the so-called CO-H$_{2}$CO-CH$_{3}$OH solid-state hydrogenation reaction chain (e.g., \citealt{Tielens82}, \citealt{Watanabe02}). Deuterated CH$_{3}$OH follows a similar formation pathway, i.e., by addition of D atoms instead of H (\citealt{WK08}, \citealt{H09}, \citealt{Taquet12b}). Thus the deuteration of methanol depends primarily on the gaseous atomic D/H ratio, which is enhanced by CO depletion (\citealt{Roberts04}). Gas-phase methanol observed towards cold cores is explained with the process of reactive desorption (\citealt{vasyunin17}), due to the exothermicity of the formation reactions on dust grain surfaces rich in CO ices. Compared to the D/H ratio which depends on the gas-phase enhancement of the main isotopolgue and is prone to the line-of-sight integration of different gas components, the D$_{2}$/D ratio provides a more stringent constraint on the deuteration fractionation in situ and is thus critical for advancing the chemical models. However, the doubly deuterated form of CH$_{3}$OH, which was first detected towards protostar IRAS 16293-2422 (\citealt{Parise02}), has not been observed towards starless and pre-stellar cores, presenting a missing link to the more evolved star-forming cores (e.g., \citealt{Dro22}, \citealt{Parise06}, \citealt{Agundez19}).

Complex organic molecules (COMs), defined as molecules with more than five atoms containing Carbon, Oxygen and Hydrogen (e.g., \citealt{Herbst09}), are ubiquitous in star-forming regions. They are very abundant around protostars, in the so-called ``hot corinos” in low-mass star forming regions, and towards ``hot massive cores” in high-mass star forming regions (e.g., \citealt{Taquet15}, \citealt{Bonfand19}). Very recently we started observing COMs also towards starless and pre-stellar cores (\citealt{Bacmann12}, \citealt{Bizzocchi14}, \citealt{JS16}), and it seems that these molecules trace very well the earliest phases of star-formation, before the ignition of the star. The formation of CH$_{3}$OH is a starting point for the molecular complexity: along the CO-H$_{2}$CO-CH$_{3}$OH formation chain, recombination of the reaction intermediates result in larger COMs (e.g., \citealt{Garrod06}, \citealt{Chuang16, Chuang17}).
However, it is still unclear how much of the chemical complexity observed towards protostars is inherited from the pre-stellar phase. Deuterium fractionation of CH$_{3}$OH is likely the best tool to provide quantitative constraints to this question.

In this paper, we report the first detection of CHD$_{2}$OH towards two pre-stellar cores, H-MM1 and L694-2.  H-MM1 is a late-stage pre-stellar core which is located in active star-forming environment Ophiuchus and characterised by strong CO depletion and high level of deuteration (\citealt{Parise11}, \citealt{Harju17}). Recent work by \citet{Pineda22} revealed evidence of NH$_{3}$ freeze-out. L694-2 is a relatively isolated pre-stellar core (\citealt{Lee11}, \citealt{Spezzano16}) and shows extended infall motions (\citealt{Keown16}, \citealt{Kim22}) resembling the proto-typical pre-stellar core L1544 (\citealt{KC10}, \citealt{Redaelli22}), but appears less centrally concentrated (\citealt{Williams06}).
We also report the non-detection of CHD$_{2}$OH towards two starless cores B68 and L1521E. Both show less CO freeze-out and smaller molecular D/H than the two pre-stellar cores (\citealt{Bergin01}, \citealt{TafallaSantiago04}, \citealt{Crapsi05}, Nagy et al, in prep.). L1521E is located in Taurus as L1544, but it appears less evolved (\citealt{Hirota02}, \citealt{TafallaSantiago04}, \citealt{Nagy19}). B68 is an isolated starless core that shows kinematic features of oscillations, presenting a stage prior to contraction (\citealt{Lada03}, \citealt{KC08}).
In this work, with the D/H ratios constrained from CH$_{2}$DOH and CHD$_{2}$OH towards the four cores, we reveal an evolutionary picture of the deuterium fractionation of CH$_{3}$OH towards low-mass cores prior to star formation. These new observations provide important constraints for the state-of-the-art chemical models, and lend support to the formation pathways of CH$_{3}$OH.

\section{Observations}\label{sec:obs}

Observations of the singly and doubly deuterated CH$_{3}$OH lines with upper level energies $E_{\mrm{up}}$$\sim$6-10 K (Table \ref{tab:lines}) were taken towards the dust peak of the target cores (Table \ref{tab:ss}) with the IRAM 30m telescope (Pico Veleta, Spain), during August to September 2022 (Project: 061-22, PI: Y. Lin) with good weather conditions ($\tau$$\sim$0.03-0.05). The EMIR E090 receiver was used with the Fourier Transform Spectrometer backend (FTS) that has a spectral resolution of 50 kHz (0.18 km s$^{-1}$ at 82.16 GHz). 
The achieved root mean square (rms) level (1$\sigma$) over the 0.18 km s$^{-1}$ channel is mostly $\sim$1-3 mK (in $T_{\mrm{mb}}$ scale), listed for each line and source specifically in Table \ref{tab:lines_obs}. The rms achieved for B68 is a bit worse, of $\sim$4 mK, as in the end less observational time was spent on the source. 
The data were processed with Gildas software\footnote{https://www.iram.fr/IRAMFR/GILDAS/} (\citealt{Pety05}). The antenna temperatures ($T_{\mathrm{A}}^{\star}$) were converted to main-beam brightness temperature ($T_{\mathrm{mb}}$) with efficiencies $B_{\mathrm{eff}}$ interpolated according to the online $\eta_{\mathrm{mb}}$ table \footnote{https://publicwiki.iram.es/Iram30mEfficiencies}, e.g., $B_{\mathrm{eff}}$ = 0.81 at 86.6 GHz.

For L1521E, we additionally adopted a previous 30m observation (PI: Z. Nagy) of the CH$_{2}$DOH 2$_{0, 2}$-1$_{0, 1}$ line. The rms level is 4 mK. The observations of the main isotopologue CH$_{3}$OH were also taken previously and published in \citet{Spezzano20}. Six $A$ and $E$-type CH$_{3}$OH lines were available. The line parameters are also listed in Table \ref{tab:lines}.

\begin{table*}
\centering
\begin{threeparttable}
\caption{Observed lines.}
\label{tab:lines}
\begin{tabular}{lccccc}
\toprule
 Molecule&Transitions  & Frequency & $E_{\mathrm{up}}$ & $A_\mathrm{ij}$&$n_{\mathrm{crit}}$\\
  &&(MHz)&(K)&10$^{-5}$ s$^{-1}$&(10$^{4}\,$cm$^{-3}$)\\
  \midrule
  CH$_{2}$DOH &1$_{1,0}$-1$_{0,1}$ $E_{0}$ & 85296.90 & 6.2 &  0.45 & - \\
   &2$_{1,1}$-2$_{0,2}$ $E_{0}$ &  86668.86 & 10.6 & 0.46 &-\\
    &$^{a}$2$_{0,2}$-1$_{0,1}$ $E_{0}$ &  89407.91 & 6.4 & 0.20 &-\\

  CHD$_{2}$OH   &2$_{0,1}$-1$_{0,1}$ $E_{0}$ & 83289.63 & 6.0 &0.22  &-\\
   &2$_{1,2}$-1$_{1,2}$ $E_{0}$ & 82165.82 & 9.1 &0.16&-\\
   \midrule
   CH$_{3}$OH &5$_{1,5}$-4$_{0,4}$ $E_{2}$&84521.17&40.4&0.20&57.1\\
&2$_{1,2}$-1$_{1,1}$ $E_{2}$&96739.35&12.5&0.26&1.3\\
&2$_{0,2}$-1$_{0,1}$ $A$&96741.37&7.0&0.34&0.7\\
&2$_{0,2}$-1$_{0,1}$ $E_{1}$&96744.54&20.0&0.34&15.4\\
&2$_{1,2}$-1$_{1,1}$ $E_{1}$&96755.50&28.0&0.26&31.5\\
&2$_{-1,2}$-1$_{-1,1}$ $A$&97582.80&21.6&0.26&59.4\\
 \bottomrule
\end{tabular}
    \begin{tablenotes}
      \small
      \item The laboratory spectroscopy references for CH$_{2}$DOH is \citet{Coudert14} (for rest frequencies) and \citealt{Pearson12}, for CHD$_{2}$OH is \citet{Dro22} and for CH$_{3}$OH \citet{Xu08}.
      \item $^{a}$ This line is only used for L1521E.
      \end{tablenotes}
  \end{threeparttable}
\end{table*}

\begin{table}
\centering
\begin{threeparttable}
\footnotesize
\caption{Information of the observed cores.}
 \label{tab:ss}
 \begin{tabular}{lllll}
\toprule
Source  & R.A. & Dec. &Type\\
   & (J2000)&(J2000)&\\
\midrule
L1521E&04:29:15.7&$+$26:14:05.0&Starless \\
B68&17:22:38.9&$-$23:49:46.0&Starless\\

H-MM1&16:27:58.3&$-$24:33:42.0&Pre-stellar   \\
L694-2&19:41:04.5&$+$10:57:02.0&Pre-stellar   \\

\bottomrule
\end{tabular}
 \begin{tablenotes}
      \small
      \item The coordinates listed correspond to the dust peak of each core (\citealt{Spezzano20}).
      \end{tablenotes}
  \end{threeparttable}
\end{table}

\section{Results}
\subsection{The obtained spectra}\label{sec:sps}
\begin{figure*}[htb]
\begin{tabular}{p{0.47\linewidth}p{0.47\linewidth}}
\hspace{0.3cm}\includegraphics[scale=0.38]{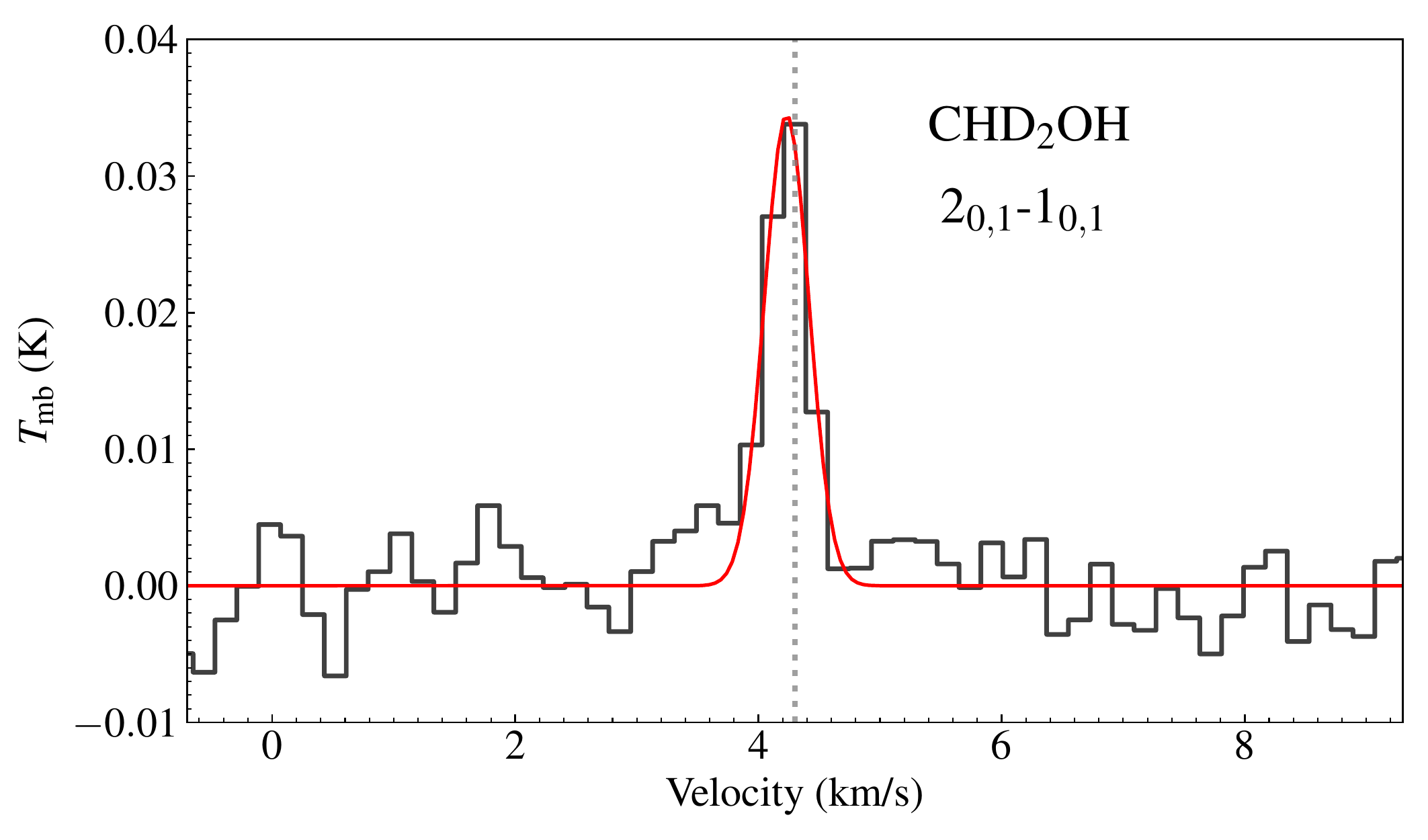}&\includegraphics[scale=0.38]{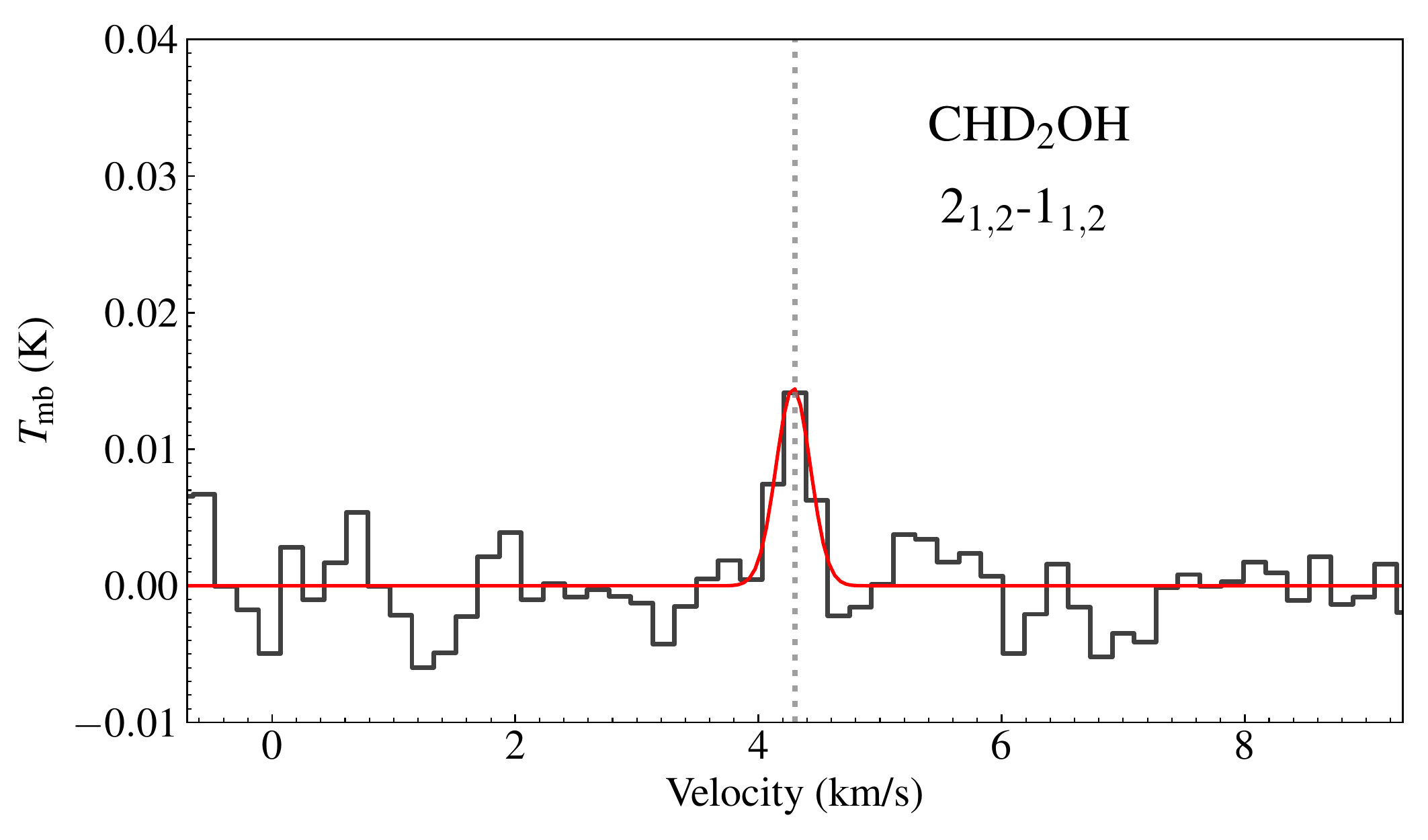}\\

\hspace{0.3cm}\includegraphics[scale=0.38]{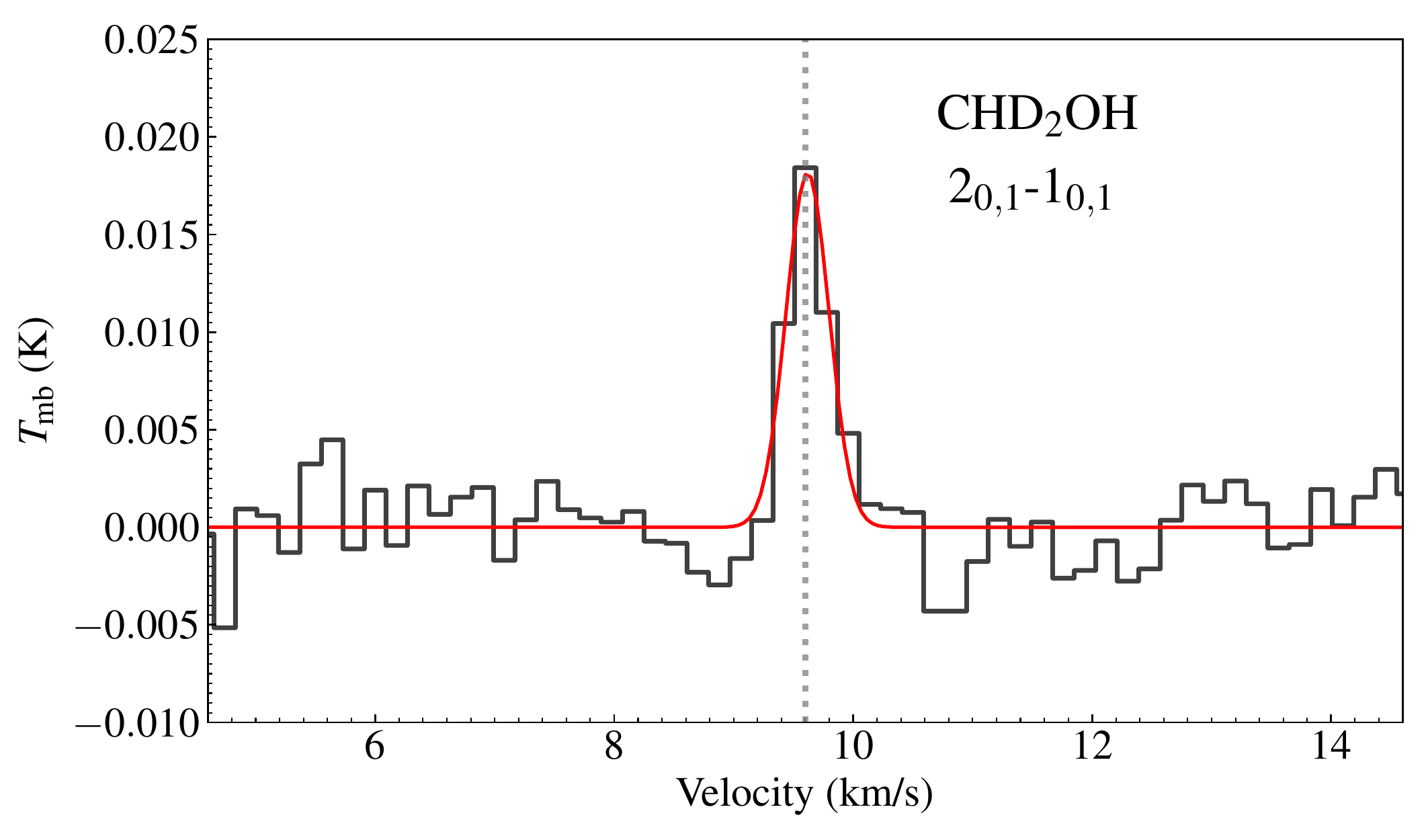}&\includegraphics[scale=0.38]{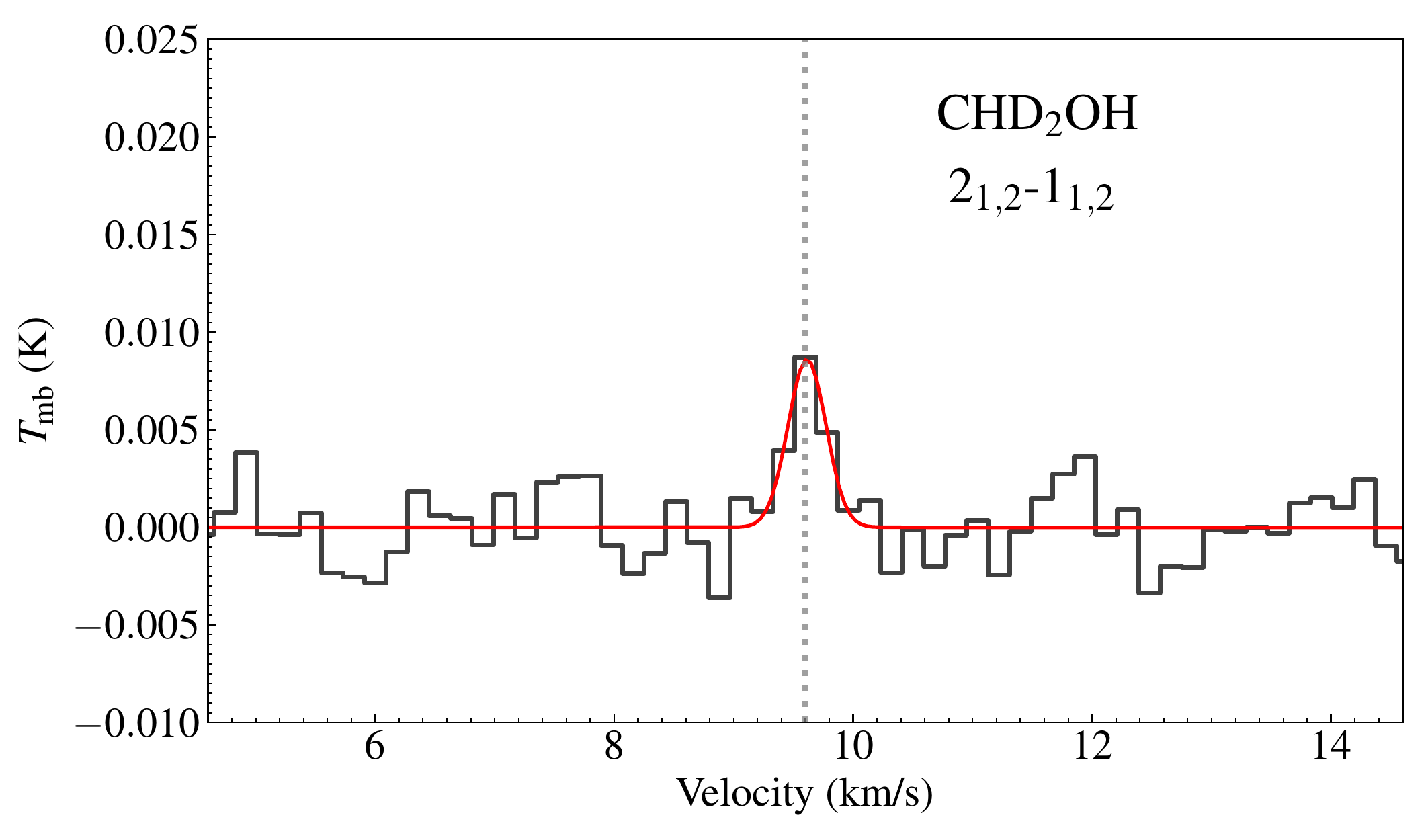}\\
\end{tabular}
\caption{CHD$_{2}$OH lines detected towards the dust peak of H-MM1 (upper panel) and L694-2 (lower panel). Vertical dotted line indicates the system velocity of CH$_{3}$OH lines. The Gaussian fit is shown as a red curve.}

\label{fig:sps_hmm1}
\end{figure*}

We detected two CH$_{2}$DOH lines and two CHD$_{2}$OH lines towards pre-stellar cores H-MM1 and L694-2. Towards the starless core L1521E we only detected CH$_{2}$DOH lines, while towards starless core B68 none of the targeted lines were detected. The two CHD$_{2}$OH lines are shown in Fig. \ref{fig:sps_hmm1}, and the two CH$_{2}$DOH lines are shown in Fig. \ref{fig:sps_app}.
The detected CH$_{2}$DOH and CHD$_{2}$OH lines were fitted with a single Gaussian profile with parameters of peak main beam brightness temperature ($T_{\mrm{mb}}$), linewidth ($\Delta v$) and system velocity ($v_{\mrm{lsr}}$). The parameters are listed in Table \ref{tab:lines_obs}. 

We find that the linewidths (Full width at half maximum, $\Delta v$) of CH$_{2}$DOH and CHD$_{2}$OH lines are similar (within 40$\%$ difference) and also close to that of CH$_{3}$OH lines. The centroid velocities ($\nu_{\mrm{LSR}}$) are also consistent between the CH$_{2}$DOH 2$_{1,1}$-2$_{0,2}$ line, two CHD$_{2}$OH lines and the CH$_{3}$OH lines. 
We note that if we use the CH$_{2}$DOH line catalogue available on the JPL database{\footnote{\url{https://spec.jpl.nasa.gov/ftp/pub/catalog/c033004.cat}}}, there is a 0.2-0.3 km/s shift between the rest velocities of the two CH$_{2}$DOH lines and the lines of the main isotopologue. This discrepancy, however, is not present in one of the CH$_{2}$DOH lines if we use the rest frequencies measured in the laboratory and tabulated in the corresponding paper (\citealt{Coudert14}). The difference between the frequencies in the catalogue and that measured in the laboratory is small but appear significant in cold sources with small linewidths, and it is the consequence of the very complex hamiltonian required to reproduce the spectra for these asymmetric internal rotors. 
We note that for cold sources with very narrow lines, the spectroscopy papers should be checked together with the catalogues for any inconsistencies.

\subsection{CH$_{3}$OH column densities}\label{sec:choh}
The observed six CH$_{3}$OH lines are listed in Table \ref{tab:lines} and span a range of $E_{\mrm{up}}$. For H-MM1 all six lines are detected (>4$\sigma$), while for other sources only the three transitions with the lowest $E_{\mrm{up}}$ are detected (2$_{1,2}$-1$_{1,1}$ $E_{2}$, 2$_{0,2}$-1$_{0,1}$ $A$, 2$_{0,2}$-1$_{0,1}$ $E_{1}$). The 3$\sigma$ noise levels were used as upper limits for the line intensities of non-detections. 

We use non-Local Thermal Equilibrium (non-LTE) RADEX (\citealt{vdt07}) models to constrain the column densities of CH$_{3}$OH ($N\mathrm{_{CH_{3}OH}}$), following the Markov chain Monte Carlo procedure elaborated in \citet{Lin22a}. The observed CH$_{3}$OH lines have a range of critical densities (Table \ref{tab:lines}), some of which are larger than the typical gas densities of pre-stellar cores, such that local thermal equilibrium (LTE) conditions are not met (c.f. \citealt{Bizzocchi14}). 
The models of $A$ and $E$-type CH$_{3}$OH lines are computed separately, considering only collisions with para-H$_{2}$ (\citealt{RF}); the ortho- to para-H$_{2}$ ratio (OPR) is assumed to be low, which is valid for dense molecular clouds, e.g., OPR$\sim$0.01 at 10$^{3}\,$cm$^{-3}$ (\citealt{Lupi21}). Collisions with ortho-H$_{2}$ are neglected in the modeling. Since the collisional de-excitation rates for temperatures lower than 10 K are not available, in the model we use the rates at 10 K for lower temperatures. The three parameters in the fitting are hydrogen volume densities ($n(\mathrm{H_{2}})$), gas kinetic temperature ($T_{\mathrm{kin}}$) and specific column densities of $A$-CH$_{3}$OH or $E$-CH$_{3}$OH ($N_{\mathrm{CH_{3}OH}}/\Delta v$). $A$ and $E$-type lines are fitted together assuming same column densities of $A$ and $E$-type CH$_{3}$OH. 

For H-MM1, $T_{\mathrm{kin}}$ and $n(\mrm{H_{2}})$ can be well constrained with six transitions, and is estimated to be 10.7$\pm$0.5 K and (1.1$\pm$0.3)$\times$10$^{6}$ cm$^{-3}$, respectively. For other sources with only three detected lines, in the fitting we assumed a $T_{\mrm{kin}}$ on average of 10 K with a standard deviation of 5 K, following a normal distribution. The $n(\mrm{H_{2}})$ for L694-2, L1521E and B68 are estimated to be 0.7$^{+1.0}_{-0.4}$$\times$10$^{4}$ cm$^{-3}$, 1.0$^{+0.5}_{-0.6}$$\times$10$^{4}$ cm$^{-3}$ and 0.5$^{+2.0}_{-0.2}$$\times$10$^{4}$ cm$^{-3}$, respectively.
As a note, the central densities of the three cores, previously estimated from dust emission, are $\sim$1-3$\times$10$^{5}$ cm$^{-3}$ (\citealt{R18}, \citealt{Tafalla04}, \citealt{Alves01}), while those for H-MM1 is $\sim$2$\times$10$^{6}$ cm$^{-3}$ (\citealt{Harju17}, \citealt{Pineda22}). Within the uncertainties, the discrepancy of the constrained $n(\mrm{H_{2}})$ with the central densities of the three cores indicates that the emission of these lower transitions of CH$_{3}$OH within the beam is dominated by outer gas layers, but nonetheless reflect the lower gas densities on average, w.r.t to H-MM1. 
The excitation temperatures $T_{\mrm{ex}}$ for the three transitions of the three cores are constrained to be $\sim$ 4-5 K, indicating sub-thermal excitations, while transitions of H-MM1 are more close to LTE. Furthermore, the model results show that the optical depths of the two transitions with lowest $E_{\mrm{up}}$ for the three sources are already moderately optically thick ($\tau$$\sim$1.0-1.5), while that of H-MM1 are optically thin ($\tau$$\sim$0.2). 
The derived total $N\mathrm{_{CH_{3}OH}}$ for the four sources are listed in Table \ref{tab:ch3oh_model}.

\begin{table}[htb]
\centering
\begin{threeparttable}
\caption{CH$_{3}$OH ($A$ and $E$) column densities derived from non-LTE models.}\label{tab:ch3oh_model}

\begin{tabular}{lllll}
\toprule
Core & \multicolumn{1}{c}{H-MM1} & \multicolumn{1}{c}{L694-2}& \multicolumn{1}{c}{L1521E}& \multicolumn{1}{c}{B68}\\
\cmidrule{2-5}
&\multicolumn{4}{c}{N/10$^{13}$ cm$^{-2}$}\\
\midrule
$N\mathrm{_{CH_{3}OH}}$&3.2(0.2)&6.0(3.0)&5.4(2.6)&3.6(1.4)\\
\bottomrule
\end{tabular}
\end{threeparttable}
\end{table}

\subsection{Singly deuterated methanol (CH$_{2}$DOH) and doubly deuterated methanol (CHD$_{2}$OH)}\label{sec:dspecies}

We calculated the column densities of CH$_{2}$DOH and CHD$_{2}$OH assuming LTE and optically thin emission, following the form (c.f. \citealt{Bizzocchi14}), 
\begin{equation*} \label{eq:N}
 N = \frac{8\pi\nu^3}{c^3} \frac{Q(T_\mrm{ex})}{g_u A_{ul}} 
     \frac{\ee^{E_u/kT_\mrm{ex}}}{\ee^{h\nu/kT_\mrm{ex}} - 1}
     \left[J_\nu(T_\mrm{ex}) - J_\nu(T_\mrm{bg})\right]^{-1}
     \int T_\mrm{mb}\diff v \,.
\end{equation*}
Here $g_{\mathrm{u}}$ is the degeneracy of the upper level, $A_{\mrm{ul}}$ is the Einstein coefficient, $E_{\mrm{up}}$ is the energy of the upper level, $T_{\mrm{ex}}$ is the excitation temperature and $Q(T_{\mrm{ex}})$ is the 
 corresponding partition function, $c$ is the light speed, $v$ is the rest frequency, $\int T_\mrm{mb}\diff v$ is the integrated emission. And $J(T)$ stands for the Rayleigh-Jeans equivalent temperature. 

The partition function for CHD$_{2}$OH is adopted from the CDMS catalog, with values updated by \citet{Dro22}. For CH$_{2}$DOH we use the JPL entry, based on \citet{Pearson12}. We further estimated the $Q(T\mathrm{_{ex}})$ at lower temperatures by summing up all energy levels ($\Sigma\,g_{\mrm{u}} e^{-E_{\mrm{up}}/k_{\mrm{B}}T_{\mrm{ex}}}$). We calculated the column densities ($N_{\mrm{CH_{2}DOH}}$ and $N_{\mrm{CHD_{2}OH}}$) with variations of $T_{\mathrm{ex}}$ within 5-8 K and assumed a calibration error of 20$\%$ for deriving the uncertainties. 
For B68 and several lines of L1521E that have no robust detection, we derived column density upper limits assuming a peak intensity of 5$\sigma$ and a linewidth of 0.5 km s$^{-1}$. 

Incorporating the possible variations of $T_{\mathrm{ex}}$ and calibration errors, the $N_{\mrm{CH_{2}DOH}}$ for H-MM1 is estimated to be 1.8$^{+0.8}_{-0.6}$$\times 10^{12}$ cm$^{-2}$, for L694-2 is 1.5$^{+0.5}_{-0.5}$$\times 10^{12}$ cm$^{-2}$. For L1521E the upper limit of $N_{\mrm{CH_{2}DOH}}$ derived from the one non-detected CH$_{2}$DOH line is 0.3$\times 10^{12}$ cm$^{-2}$, compatible with the value estimated from one of the detected lines CH$_{2}$DOH 1$_{1,1}$-1$_{0,1}$, but not the other, the 2$_{0,2}$-1$_{0,1}$ line. This is possibly due to the higher noise level of the latter observation of $\sim$3.5 mK (compared to 1.5 mK). Therefore for L1521E we took $N_{\mrm{CH_{2}DOH}}$ as 0.3$^{+0.2}_{-0.1}$$\times 10^{12}$ cm$^{-2}$, accounting for these uncertainties. Upper limit of $N_{\mrm{CH_{2}DOH}}$ of B68 is 0.5$\times$10$^{12}$ cm$^{-2}$. 
The $N_{\mrm{CHD_{2}OH}}$ for H-MM1 and L694-2 are 1.4$^{+0.5}_{-0.4}$$\times 10^{12}$ cm$^{-2}$ and 0.8$^{+0.3}_{-0.2}$$\times 10^{12}$ cm$^{-2}$, respectively. Upper limits of $N_{\mrm{CHD_{2}OH}}$ for B68 and L1521E are 1.0$\times 10^{12}$ cm$^{-2}$ and 0.4$\times 10^{12}$ cm$^{-2}$. The difference of the upper limits originates from the different rms levels reached for the two sources. The calculated column densities are listed in Table \ref{tab:NCH2DOH}.





\begin{table*}[htb]
\centering
\begin{threeparttable}
\caption{CH$_{2}$DOH and CHD$_{2}$OH column densities calculated with LTE and optically thin assumptions using different excitation temperatures.}\label{tab:NCH2DOH}
\footnotesize
\begin{tabular}{lllllll}
\toprule

&\multicolumn{6}{c}{N/10$^{12}$ cm$^{-2}$$^{a}$} \\\cmidrule{2-7}
$T_{\mathrm{ex}}$ = &5 K &  6.5 K &               8 K&             5 K &    6.5 K  &  8 K \\
\midrule
Core & \multicolumn{3}{c}{H-MM1} & \multicolumn{3}{c}{L694-2}\\
\midrule

CH$_{2}$DOH 1$_{1, 1}$-1$_{0, 1}$ &1.6$\pm$0.4 &1.6$\pm$0.3&1.8$\pm$0.4&1.3$\pm$0.3&1.3$\pm$0.3&1.5$\pm$0.3\\
CH$_{2}$DOH 2$_{1, 1}$-2$_{0, 2}$    &2.1$\pm$0.5&1.8$\pm$0.3 &1.8$\pm$0.4&1.7$\pm$0.3&1.4$\pm$0.4&1.4$\pm$0.3\\

CHD$_{2}$OH 2$_{0, 1}$-1$_{0, 1}$      &1.3$\pm$0.2&1.2$\pm$0.2&1.2$\pm$0.2&0.7$\pm$0.1&0.8$\pm$0.2&0.9$\pm$0.2\\
CHD$_{2}$OH 2$_{1, 2}$-1$_{1, 2}$        &1.2$\pm$0.2&1.4$\pm$0.3&1.6$\pm$0.3&0.7$\pm$0.1&0.7$\pm$0.1&0.7$\pm$0.1\\
\midrule
 & \multicolumn{3}{c}{B68} & \multicolumn{3}{c}{L1521E}\\
\midrule
CH$_{2}$DOH 1$_{1, 1}$-1$_{0, 1}$ &<0.5 &<0.6&<0.6&0.2$\pm$0.1&0.2$\pm$0.1&0.2$\pm$0.1\\
CH$_{2}$DOH 2$_{1, 1}$-2$_{0, 2}$    &<0.8&<0.7&<0.7&<0.3&<0.3&<0.3\\
CH$_{2}$DOH 2$_{0, 2}$-1$_{0, 1}$ &-&-&-&0.5$\pm$0.1&0.5$\pm$0.1& 0.6$\pm$0.1\\
CHD$_{2}$OH 2$_{0, 1}$-1$_{0, 1}$      &<2.0&<2.0&<2.0&<0.7&<0.7&<0.7\\
CHD$_{2}$OH 2$_{1, 2}$-1$_{1, 2}$        &<0.8&<0.9&<1.0&<0.3&<0.4&<0.4\\
\bottomrule
\end{tabular}
\begin{tablenotes}
\item $^{a}$ Uncertainties are calculated considering 20$\%$ calibration errors.
\end{tablenotes}
\end{threeparttable}
\end{table*}




             

\subsection{Discussion and conclusions}
We can first put the deuteration of CH$_{3}$OH into a general context of molecular deuteration in pre-stellar cores. Deuterations of various molecules are well studied towards the proto-typical pre-stellar core L1544, which is on the verge of protostar formation and has a central density of $\sim$10$^{7}$\,cm$^{-3}$ (\citealt{KC10}). The two pre-stellar cores in our sample, L694-2 and H-MM1 appear less dense in the center than L1544 (see Sect. \ref{sec:choh}).
Towards the dust peak of L1544, the column density ratios of the singly deuterated molecules w.r.t to the main isotopomers are [$c$-C$_{3}$HD]/[$c$-C$_{3}$H$_{2}$]$\sim$0.17, [HDCO]/[$\mrm{H_{2}CO}$]$\sim$0.04, [HDCS]/[$\mrm{H_{2}CS}$]$\sim$0.12, $[\mrm{DCO^+}]$/$[\mrm{HCO^+}]$$\sim$0.04, $[\mrm{N_{2}D^+}]$/$[\mrm{N_{2}H^+}]$$\sim$0.26 (\citealt{CT19}, \citealt{Redaelli19}, \citealt{Spezzano22}). For CH$_{3}$OH, \citet{Bizzocchi14} and \citet{CT19} derived a $[\mrm{CH_{2}DOH}]/[\mrm{CH_{3}OH}]$$\sim$0.08, which shows that the deuteration level of CH$_{3}$OH is intermediate among these molecules. 
\citet{Vastel14} suggests that the emission of CH$_{3}$OH towards L1544 in fact traces an outer layer of the core (see also \citealt{Spezzano17}, \citealt{Lin22b}).
The ratios of the column densities of doubly deuterated molecules and the singly deuterated ones, however, show a different picture among these molecules, with 
[$c$-$\mrm{C_{3}D_{2}}$]/[$c$-$\mrm{C_{3}HD}$]$\sim$0.10, [$\mrm{D_{2}CO}$]/[HDCO]$\sim$1.15, [HDCS]/[$\mrm{H_{2}CS}$]$\sim$1.00 (\citealt{Spezzano13, Spezzano22}). 

Since deuteration fractionation depends heavily on the physical and chemical parameters of the cores, it is critical to verify if the observed D/H ratios of the dense cores are legacies from the preceding evolutionary stages and remain as fossil records, and how they differ with that in comets and meteorites of solar system (\citealt{CC14}, \citealt{Nomura22}). These questions can be answered by examining the variations of D/H ratios in dense cores prior to star formation. In this context, \citealt{Chantzos18} find that the D/H ratios of $c$-C$_{3}$H$_{2}$ do not show significant variations over the core evolution. On the other hand, the D/H ratios of N$_{2}$H$^+$ show a considerable increase in the dynamically more evolved pre-stellar core L1544 (\citealt{Crapsi05}). As a note, the N$_{2}$D$^+$/N$_{2}$H$^{+}$ ratio of H-MM1 is $\sim$0.43 (\citealt{Punanova16}), and N$_{2}$D$^+$/N$_{2}$H$^{+}$ of L1544 and L694-2 are $\sim$0.25 (\citealt{Crapsi05}).
The D/H ratios of HCO$^{+}$ and H$_{2}$CO are suggested to have a similar evolutionary effect, showing correlations with CO depletion and gas density (\citealt{Caselli99}, \citealt{Bacmann02, Bacmann03}).

\begin{figure*}[htb]
\begin{tabular}{p{0.85\linewidth}}
\hspace{1cm}\includegraphics[scale=0.32]{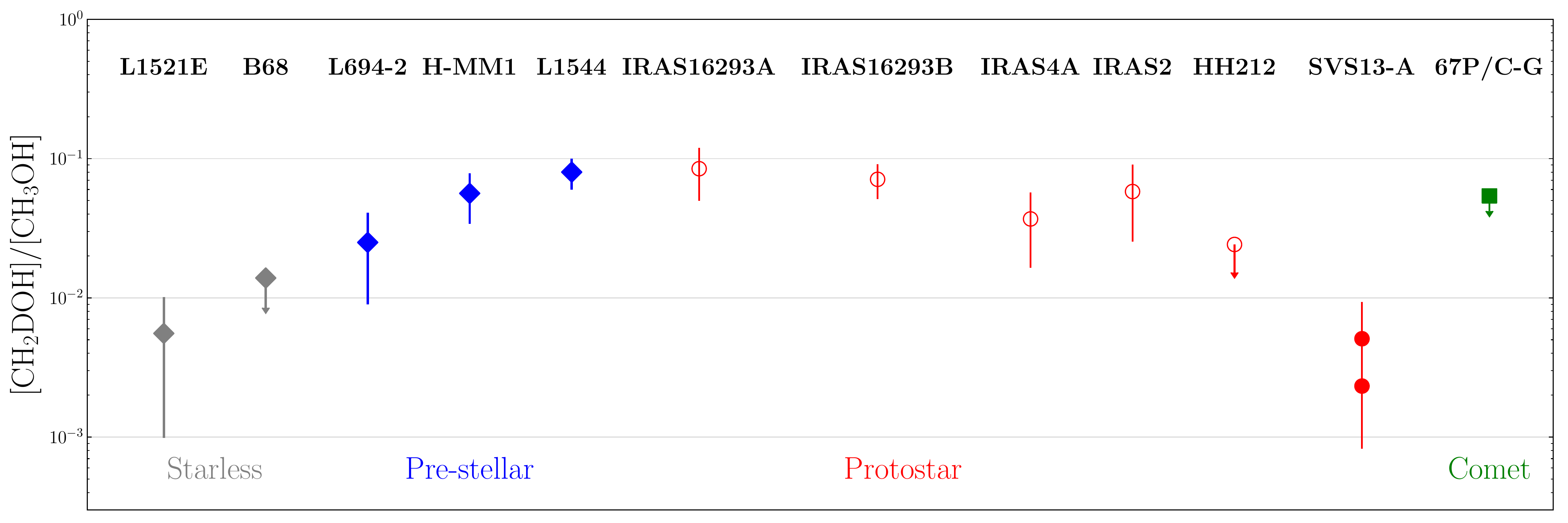}\\
\hspace{1cm}\includegraphics[scale=0.32]{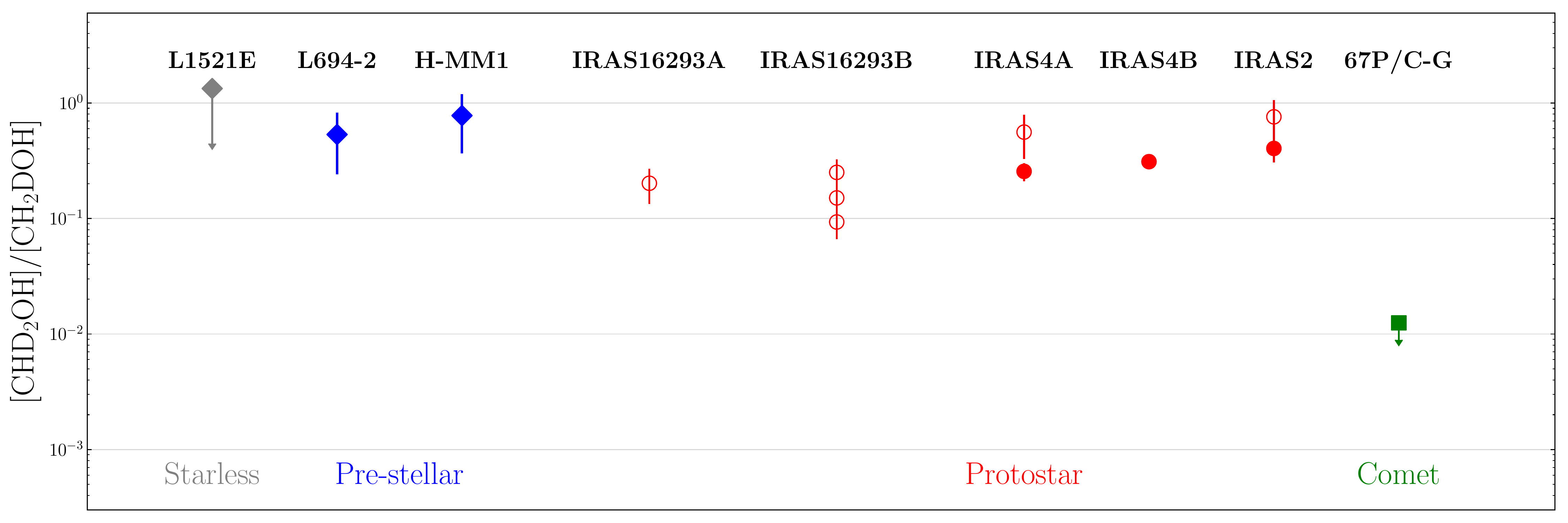}\\
\end{tabular}
\caption{The column density ratios of $[\mrm{CH_{2}DOH}]/[\mrm{CH_{3}OH}]$ ({\it{upper panel}}) and $[\mrm{CHD_{2}OH}]/[\mrm{CH_{2}DOH}]$ ({\it{lower panel}}) as a function of source types. Filled markers indicate single-dish observations and open markers indicate interferometric observations. Markers with downward arrows show the derived upper limits. The references for SVS13-A is \citet{Bianchi17a}; for IRAS4A and IRAS2 are \citet{Taquet19} and \citet{Parise06} (also IRAS4B, all located in NGC 1333); for HH212 is \citet{Bianchi17b}; for IRAS16293A and IRAS16293B are \citet{Manigand19}, \citealt{Jorgensen16}, \citet{Dro22}; for comet 67P/C-G is \citet{Dro21}. For an exhaustive plot of $[\mrm{CH_{2}DOH}]/[\mrm{CH_{3}OH}]$ of all available measurements we refer to \citet{Dro21} and their supplementary data.}
\label{fig:ab_sum}
\end{figure*}

We estimated the column density ratios of [$\mrm{CH_{2}DOH}$]/[$\mrm{CH_{3}OH}$] and [$\mrm{CHD_{2}OH}$]/[$\mrm{CH_{3}OH}$] for the four starless and pre-stellar cores. 
We present summary plots in Fig. \ref{fig:ab_sum} of four types of objects: starless and pre-stellar cores, protostars, and comets, with values collected from the literature.
The $[\mrm{CH_{2}DOH}]/[\mrm{CH_{3}OH}]$ for H-MM1 and L694-2 is 0.06$\pm$0.02 and 0.03$\pm$0.02, respectively, which are lower than what is measured towards L1544 (\citet{Bizzocchi14}, \citealt{CT19}); starless cores L1521E and B68 show considerably lower $[\mrm{CH_{2}DOH}]/[\mrm{CH_{3}OH}]$ values, for L1521E the ratio is 0.006$\pm$0.005, and the upper limit for B68 is 0.014 (Fig. \ref{fig:ab_sum}, left panel). 
Meanwhile, the $[\mrm{CHD_{2}OH}]/[\mrm{CH_{2}DOH}]$ for H-MM1 and L694-2 are 0.8$\pm$0.4 and 0.5$\pm$0.3. Compared with $[\mrm{CH_{2}DOH}]/[\mrm{CH_{3}OH}]$, these ratios suggest a higher D/H ratio of the di-deuterated molecules than the mono-deuterated ones, showing a similar picture as that in Class 0 objects (\citealt{Dro22}). High-angular resolution, interferometric observations towards Class 0 objects reveal that the $[\mrm{CHD_{2}OH}]/[\mrm{CH_{2}DOH}]$ are $\sim$0.2-0.7 (\citealt{Dro22}, \citealt{Taquet19}), consistent with our results based on single-dish observations towards pre-stellar cores (Fig. \ref{fig:ab_sum}, right panel). 
These higher than 0.5 ratios of $[\mrm{CHD_{2}OH}]/[\mrm{CH_{2}DOH}]$ are also consistent those derived by single-dish observations towards hot corinos (\citealt{Parise06}, \citealt{Agundez19}, \citealt{Bianchi17a}), with the latter supposedly tracing more of the extended envelope. Since the deuteration in hot corino is expected to be lower than in the colder envelope due to the gradient of deuteration in ices and evaporations of different parts of the ices (\citealt{Taquet14}), our results, which are limited by angular resolution, do not yet suggest a strong evolutionary effect of $[\mrm{CHD_{2}OH}]/[\mrm{CH_{2}DOH}]$ from the pre-stellar stage to the star-forming stage (Fig. \ref{fig:ab_sum}, right panel). Future interferometric observations of CHD$_{2}$OH towards pre-stellar cores are required to properly separate different emitting gas components (that of the main isotopologue as well), and to pin down the emission region for more stringent determinations of the column density ratios. 

Taking into consideration the statistical correction associated with the CH$_{3}$ functional group (c.f. Appendix B of \citealt{Manigand19}), the D/H ratios calculated from $[\mrm{CH_{2}DOH}]/[\mrm{CH_{3}OH}]$ and $[\mrm{CHD_{2}OH}]/[\mrm{CH_{3}OH}]$ for the two pre-stellar cores range from 0.01-0.02 and 0.07-0.12 (Table \ref{tab:ratios}), respectively, consistent with that towards Class 0 objects (\citealt{Parise06}, \citealt{Taquet19}, \citealt{Dro22}). While starless cores L1521E and B68 show a lower D/H value (or upper limits). The increasing and relatively high D/H ratios of CH$_{3}$OH in more evolved pre-stellar cores is a strong observational evidence that methanol form in relatively late-stage cores which are prior to collapse, an environment characterised by high D/H atomic ratio and abundant CO in the ice for hydrogenation.  
We summarise the two D/H
ratios of the four cores in Table \ref{tab:ratios}.

\begin{table}[htb]
\centering
\begin{threeparttable}
\caption{D/H ratios derived from abundance ratios of $[\mrm{CH_{2}DOH}]/[\mrm{CH_{3}OH}]$ and $[\mrm{CHD_{2}OH}]/[\mrm{CH_{3}OH}]$ considering statistical corrections.}\label{tab:ratios}

\begin{tabular}{lllll}
\toprule
Used abundance ratio &\multicolumn{4}{c}{Derived D/H ($\%$)}\\
 & \multicolumn{1}{c}{H-MM1} &\multicolumn{1}{c}{L694-2}&\multicolumn{1}{c}{L1521E}&\multicolumn{1}{c}{B68}\\
\midrule
CH$_{2}$DOH/CH$_{3}$OH &1.9$\pm$0.7&0.8$\pm$0.5&0.2$\pm$0.1&<0.5\\
CHD$_{2}$OH/CH$_{3}$OH &12.1$\pm$2.2&7.0$\pm$2.1&<5&<9\\
\bottomrule
\end{tabular}
\begin{tablenotes}
\small
\item According to Table B. 1 of \citet{Manigand19}, the relations of the molecular abundance ratio with D/H ratio are CH$_{2}$DOH/CH$_{3}$OH = 3(D/H) and CHD$_{2}$OH/CH$_{3}$OH = 3(D/H)$^{2}$.  
\item Upper limits are calculated based on 5$\sigma$ noise levels of CH$_{2}$DOH and CHD$_{2}$OH lines (Table \ref{tab:lines_obs}).
\end{tablenotes}
\end{threeparttable}
\end{table}

\citet{Ambrose21} observed CH$_{2}$DOH towards 12 starless and pre-stellar cores in Taurus and revealed an evolutionary variation of the D/H ratio. Our new observations confirm the evolutionary trend of deuterium fraction of CH$_{3}$OH, from both singly- and doubly-deuterated forms, for starless and pre-stellar cores (Table \ref{tab:ratios}). As previously suggested, the extreme deuterium fraction of the more evolved, star-forming cores is likely inherited from the pre-stellar stage (\citealt{CC14}). Linking to the measurements towards the volatiles and the dust of comet, \citet{Dro21} revealed a similar D/H ratio estimated from CH$_{2}$DOH of comet 67P/Churyumov-Gerasimenko (hereafter 67P/C-G) and Class 0/I objects; we confirmed that the chemical heritage is generated from the pre-stellar stage. On the other hand, D/H based on CHD$_{2}$OH show higher levels than comet 67P/C-G, which may imply a further processing of the doubly deuterated form or small-scale non-uniformity down to the comet scale. In particular, the warming up of the originally cold extended envelope, which has a lower molecular D/H ratio, likely causes the decrease of the observed molecular D/H from Class 0 to Class I object (\citealt{Bianchi17b}, see Fig. \ref{fig:ab_sum} for SVS-13A, the only Class I object in category Protostar). There might be a similar origin for the materials ending up in comet; further evidence may be collected from more observations on the D/H ratios in spatially resolved Class I objects. Another possibility to explain this difference is that our Solar System probably formed in a warmer environment, e.g., in a large cluster-forming region, where external illumination may have increased the overall temperature of the Solar Nebula (\citealt{Adams10}), and is characterised by lower D/H ratio.

Compared to the laboratory experiment results of the D/H (\citealt{N05, N07}, \citealt{H09}), D/H from CHD$_{2}$OH towards pre-stellar cores are lower, while D/H from CH$_{2}$DOH are consistent with the measurements. 
\citealt{Dro22} attribute this discrepancy as the over H-D substitution for the formation of the di-deuterated species; the over-availability of D atoms in the experiments resembles more of the conditions in pre-stellar stage. Our results confirm that towards the pre-stellar cores, the inconsistencies with experimental predictions still persist. On the other hand, the high D$_{2}$/D ratio towards pre-stellar cores may point to an efficient H addition to D$_{2}$CO in the formation of CHD$_{2}$OH (\citealt{H09}). In fact, the D$_{2}$CO/HDCO column density ratio towards one proto-stellar core SM1 in $\rho$ Ophiuchi A (\citealt{Kawabe18}, \citealt{Chen18}), is estimated to be larger than 1 (\citealt{Bergman11}), indicative of the efficient abstraction and substitution process in forming D$_{2}$CO as ingredients of CHD$_{2}$OH. Similarly, the D$_{2}$CS/HDCS ratio was also found to be $\sim$1 towards L1544 (\citealt{Spezzano22}), suggesting an efficient formation of the doubly deuterated form D$_{2}$CX (with X = O or S). 

Comparing the observed D/H with current theoretical studies of methanol deuteration based on gas-grain chemical models (\citealt{Taquet14}), the D/H from CH$_{2}$DOH are compatible, while D/H from CHD$_{2}$OH is strongly underestimated (a factor of $\sim$10) by the models. This, together with the discrepancy between the observed D/H from CHD$_{2}$OH and the aforementioned laboratory results, suggest that our current understanding of the formation pathways of the doubly deuterated CH$_{3}$OH, and in a broader sense, the modeling of chemical reactions on interstellar ices, still need to be improved in future works.


\begin{acknowledgements}
      The authors acknowledge the financial support of the Max Planck Society. YL wishes to thank Pablo Torne, Damour Fr{\'e}d{\'e}ric and Santiago Joaq{\'i}n for their kind help during the 30m observations.
      \end{acknowledgements}

\bibliography{ref}

\begin{appendix}

\section{CH$_{2}$DOH spectra}
\begin{figure*}[htb]
\begin{tabular}{p{0.47\linewidth}p{0.47\linewidth}}
\centering\hspace{0.3cm}\includegraphics[scale=0.38]{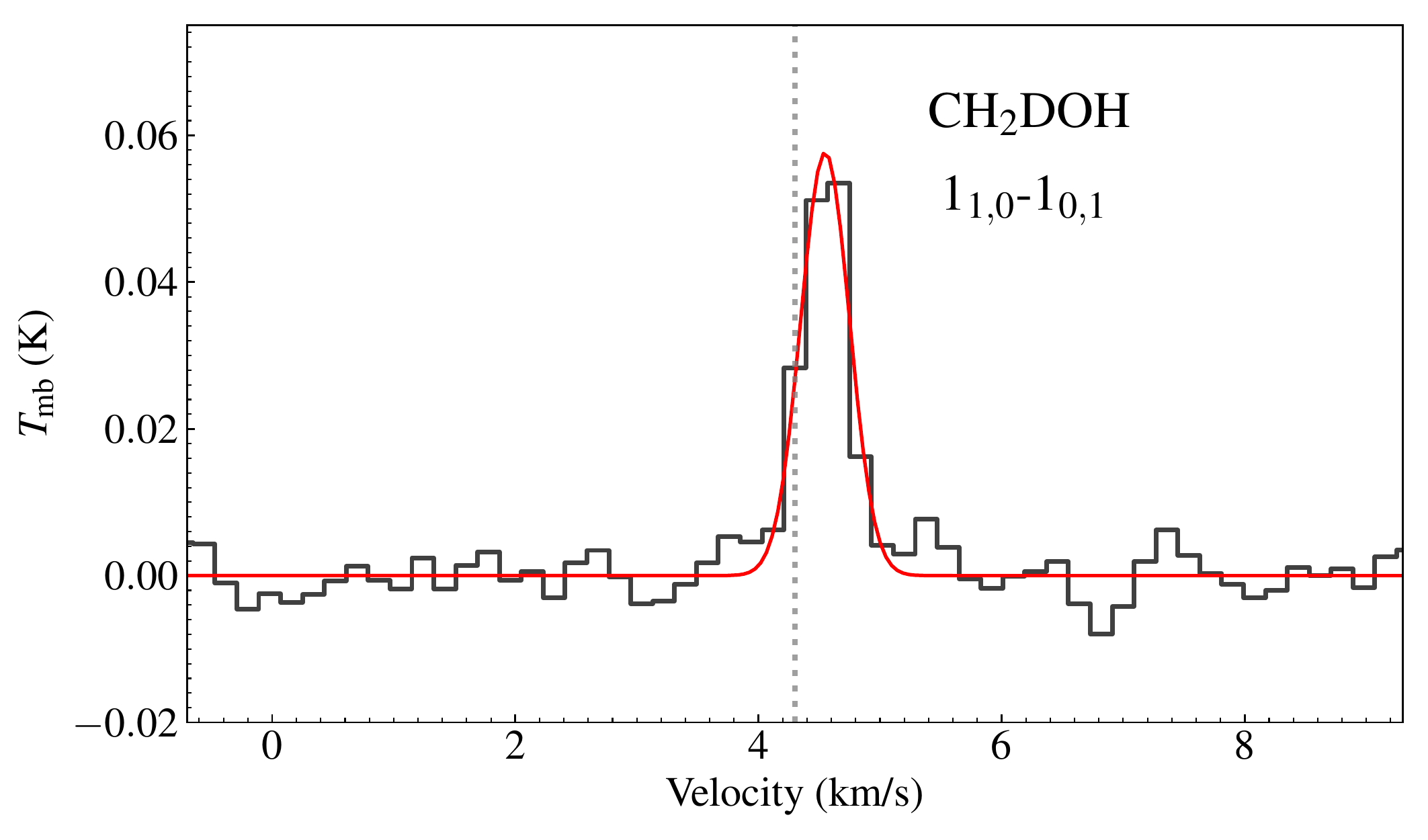}&\includegraphics[scale=0.38]{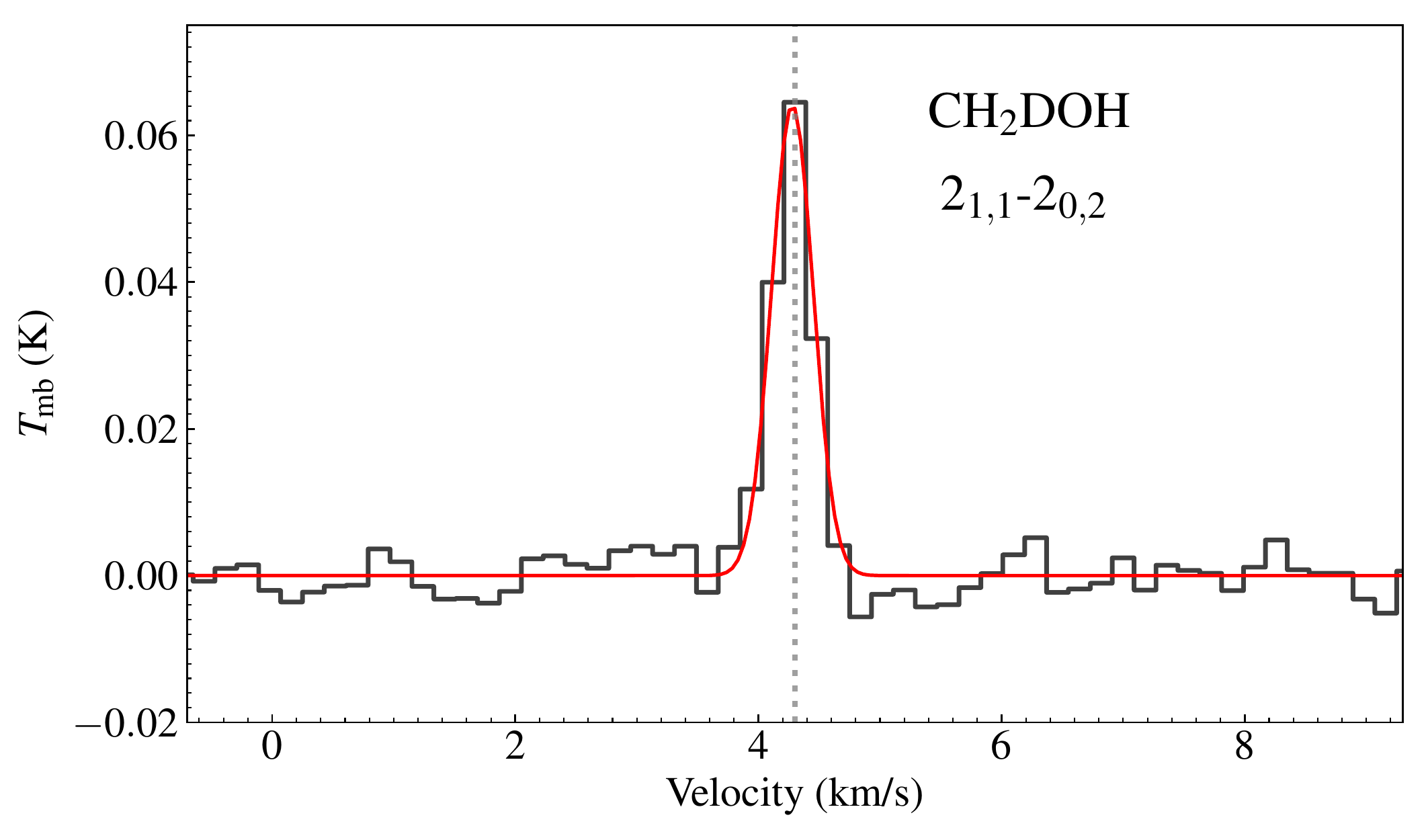}\\
\hspace{0.3cm}\includegraphics[scale=0.38]{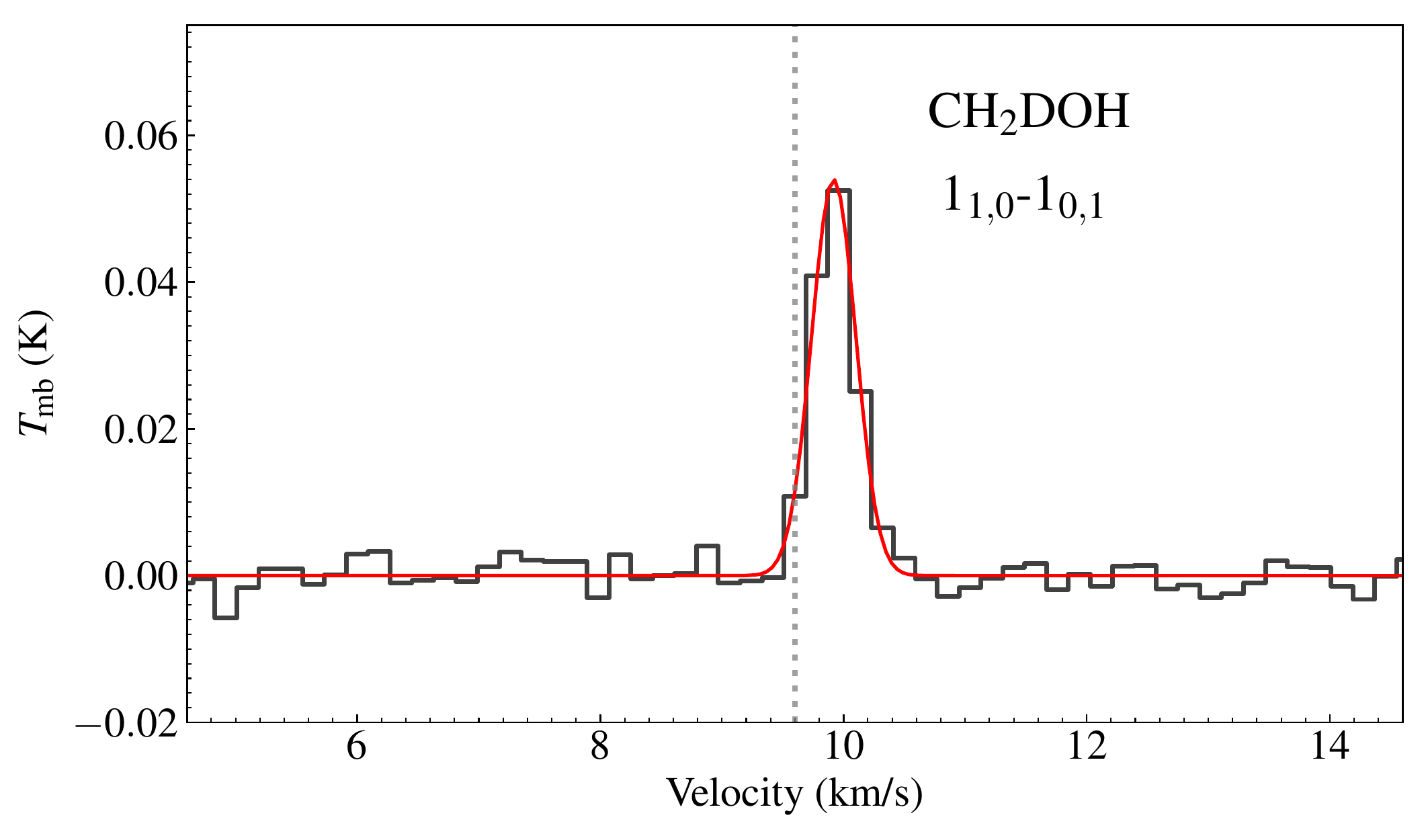}&\includegraphics[scale=0.38]{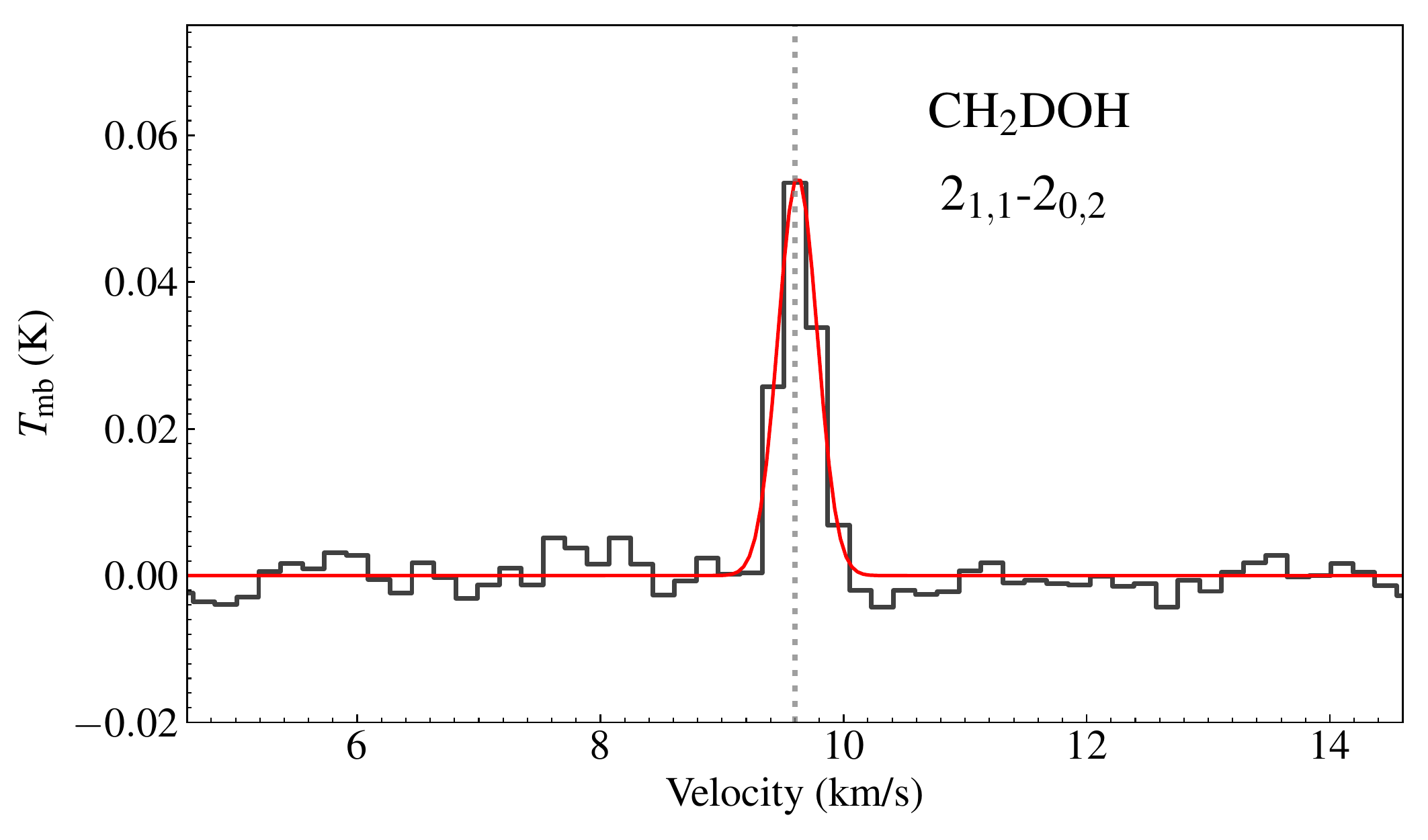}\\
\end{tabular}
\caption{CH$_{2}$DOH lines detected towards dust peak of H-MM1 (upper panel) and L694-2 (lower panel). Vertical dotted line indicates the system velocity of CH$_{3}$OH lines. The Gaussian fit is shown as a red curve.}

\label{fig:sps_app}
\end{figure*}

The two CH$_{2}$DOH spectra observed towards H-MM1 and L694-2 are shown in Fig. \ref{fig:sps_app}.

\section{Line parameters}
The line parameters of our obtained spectra towards the four cores are listed in Table \ref{tab:lines_obs}.
\begin{table*}[htb]
\centering
\begin{threeparttable}
\caption{Line parameters of CH$_{2}$DOH and CHD$_2$OH of the cores.}\label{tab:lines_obs}

\begin{tabular}{lllllllllll}
\toprule

Line parameter &  $\int T_{\mathrm{mb}}\,dv$ &                 $\nu_{\mathrm{LSR}}$ &              $\Delta v$ & rms&SNR&$\int T_{\mathrm{mb}}\,dv$ &     $\nu_{\mathrm{LSR}}$ &  $\Delta v$ &rms&SNR\\
              &   mK km s$^{-1}$   &         km s$^{-1}$          &    km s$^{-1}$             &       mK      &  & mK km s$^{-1}$&    km s$^{-1}$              &    km s$^{-1}$  &mK &    \\\cmidrule{2-6}\cmidrule{7-11}
Core & \multicolumn{5}{c}{H-MM1} & \multicolumn{5}{c}{L694-2}\\
\midrule
CH$_{2}$DOH 1$_{1, 0}$-1$_{0, 1}$    &28.8(1.4) & 4.5(0.01)   &0.49(0.03)    &    2.5               &  22     &24.1(0.7)      &        9.9(0.01)&   0.37(0.01)&  1.9&32 \\
CH$_{2}$DOH 2$_{1, 1}$-2$_{0, 2}$     &   27.2(1.1)  & 4.3(0.01)   &0.42(0.02)  &1.9             &  32    &    21.7(0.8)    &        9.6(0.01)&  0.39(0.02) & 1.9&27 \\
CHD$_{2}$OH 2$_{0, 1}$-1$_{0, 1}$   &15.8(1.3)&4.2(0.02)&0.43(0.04) &2.1 &16&8.6(0.8)            &  9.6(0.02)           &      0.43(0.05)    & 1.5  &12   \\
CHD$_{2}$OH 2$_{1, 2}$-1$_{1, 2}$   &5.9(1.1)&4.3(0.03)&0.36(0.07)&2.4  &    6    &3.4(0.7)   & 9.6(0.04)   &0.38(0.10)      &1.7    &5   \\\cmidrule{2-5}\cmidrule{6-11}

&\multicolumn{5}{c}{L1521E} & \multicolumn{5}{c}{B68}\\
\midrule
CH$_{2}$DOH 1$_{1, 0}$-1$_{0, 1}$    &2.9(0.5) & 6.5(0.01)   &0.39(0.11)    &    1.3              &5&<10.0&-&-&4.0&-\\
CH$_{2}$DOH 2$_{1, 1}$-2$_{0, 2}$ &<4.0&-&-&1.5&-&<10.0&-&-&4.0&-\\
CH$_{2}$DOH 2$_{0, 2}$-1$_{0, 1}$     &   6.1(1.0)  & 6.7(0.05)   &0.41(0.09)  &3.5        &4&-&-&-&-&- \\
CHD$_{2}$OH 2$_{0, 1}$-1$_{0, 1}$   &<4.0&-&-&1.5&-&<10.0&-&-&4.0&-\\
CHD$_{2}$OH 2$_{1, 2}$-1$_{1, 2}$   &<4.0&-&-&1.5&-&<10.0&-&-&4.0&-\\
\bottomrule
\end{tabular}
\begin{tablenotes}
\item - indicates non-detection and the corresponding upper limit of integrated intensity is estimated assuming 5$\sigma$ with a linewidth $\Delta v$ of 0.5 km s$^{-1}$.
\end{tablenotes}
\end{threeparttable}
\end{table*}

\section{The [$\mrm{CHD_{2}OH}$]/[$\mrm{CH_{3}OH}$] ratios}
Similar to Fig. \ref{fig:ab_sum}, we plot the [$\mrm{CHD_{2}OH}$]/[$\mrm{CH_{3}OH}$] ratios as a function of source types. 

\begin{figure*}[htb]
\hspace{1cm}\includegraphics[scale=0.32]{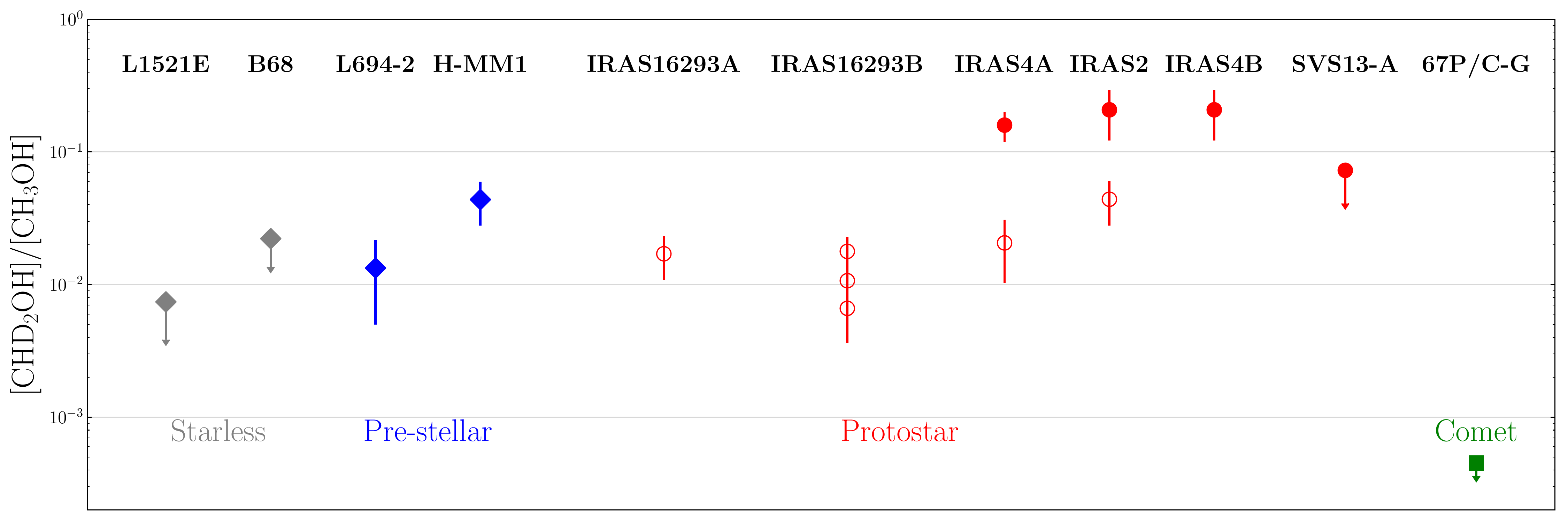}
\caption{Same as Fig. \ref{fig:ab_sum}, but for the [$\mrm{CHD_{2}OH}$]/[$\mrm{CH_{3}OH}$] ratios. }
\end{figure*}

\end{appendix}
\end{document}